\documentstyle[11pt]{article}
\input epsf
\begin{document}
\baselineskip 18pt
\vskip 5mm
\centerline{\Large \bf
NONEQUILIBRIUM QUANTUM SCALAR FIELDS
IN COSMOLOGY\footnote{To appear in the Proceedings of the
Fourth International Workshop on Thermal
Field Theories and Their Applications,
Dalian China, 1995}}
\vskip 5mm
\centerline{\sc Sang Pyo Kim}
\vskip 5mm
\centerline{\it Department of Physics,
Kunsan National University}
\centerline{\it
Kunsan 573-701, Korea}
\vskip 5mm
\begin{quote}
We elaborate further the functional
Schr\"{o}dinger-picture approach to the quantum field
in curved spacetimes using the generalized invariant
method
and construct explicitly the Fock space, which
we relate with the thermal field theory.
We apply the method to a free massive scalar field
in the de Sitter spacetime, and find the exact
quantum states, construct the Fock space,
and evaluate the two-point function
and correlation function.
\end{quote}
\section{Introduction}

Quantum field theory in curved spacetimes (QFTCS)
has been a hot issue revisited recently and applied
widely to cosmology.
There have been two conventional canonical
approaches to QFTCS: one is based on the solution
space of classical equations \cite{davis} and
the other is the functional
Schr\"{o}dinger-picture field theory
\cite{jackiw}. One may show that
these two canonical methods give the identical results
in the end. The other nonconventional
approach is the semiclassical gravity,
a kind of QFTCS, in which not only the quantum
back reaction of matter field but also
the quantum gravitational corrections to matter field
are considered and has
been applied successfully and usefully to cosmology,
derived from the Wheeler-DeWitt equation
(see \cite{kim} for a self-consistent method
and recent references).
But this does not imply that
the canonical quantum gravity
based on the Wheeler-DeWitt equation
is free of all the conceptual and
technical problems.

If quantum gravity written formally in the form
$\hat{G}_{\mu\nu} = 8\pi \hat{T}_{\mu\nu}$ succeeds
indeed, there should be some limiting procedures from
quantum gravity to semiclassical
gravity written formally in the form
$G_{\mu\nu} = 8\pi \left<\hat{T}_{\mu\nu} \right>$, in which
the background gravity treated as classical
is equated with the expectation value of
quantum energy-momentum tensor. It should be remarked that
regardless of approaches
QFTCS is characterized by time-dependence due to the
time-dependent metrics of spacetimes in general, that is,
quantum fields are nonequilibrium or out-of-equilibrium.

In this talk we shall elaborate the
functional Schr\"{o}dinger-picture approach
further by unifying
the generalized invariant method with the functional
Schr\"{o}dinger-picture approach to quantum field
theory in curved spacetimes.
One of the motivations
for introducing the functional
Schr\"{o}dinger-picture approach was
its effectiveness  and usefulness
in treating these time-dependent Hamiltonian systems
of fields.
On the other hand, the generalized invariant method
has been used
to find explicitly the exact quantum states of
time-dependent Hamiltonian systems in quantum mechanics
\cite{lewis}.
Although these two
methods were introduced largely for time-dependent
quantum field theory and quantum mechanical systems,
there has not been yet any attempt to combine these two
approaches in particular to a field theoretical
framework. The method introduced in this talk
applies equally to not only the conventional canonical approach
in functional Schr\"{o}dinger-picture but also
the semiclassical gravity from the Wheeler-DeWitt equation.

\section{Functional Schr\"{o}dinger Equation}

To simplify the model under study,
we shall consider a free massive
scalar field in a spatially flat ($ k =1$)
Friedmann-Robertson-Walker universe with the metric
\begin{equation}
ds^2 = - dt^2 + R^2 (t) d{\rm x}^2.
\end{equation}
The Hamiltonian for the free massive scalar field
takes the form
\begin{equation}
H(t) = \int d{\rm x}  \left[
\frac{1}{2}\frac{1}{R^3(t)} \pi^2({\rm x}, t)
+ \frac{1}{2} R(t) (\nabla \phi({\rm x}, t))^2
+ \frac{1}{2} R^3(t) \left( m^2 + \xi
{\bf R} \right) \phi^2 ({\rm x},t) \right],
\end{equation}
where ${\bf R}$ is the three-curvature.
$\xi = 0$ is the minimal scalar field and
$\xi = 1/6$ is the conformal scalar field.
We decompose the scalar field into modes by
confining it into a box of size $l$
and use the orthonormal basis
\begin{equation}
u_{\bf n}^{(+)} ({\bf x})
= \sqrt{2/l^3} \cos\left(2\pi {\bf n}
\cdot {\bf x} / l\right),
u_{\bf n}^{(-)} ({\bf x})
= \sqrt{2/l^3} \sin\left(2\pi {\bf n}
\cdot {\bf x} / l\right).
\end{equation}
In a compact notation, $\sum_{\pm, {\bf n}} =
\sum_{\alpha}$, the scalar field and its momentum
can be expanded as
\begin{equation}
\phi ({\bf x}, t) = \sum_{\alpha}
\phi_\alpha (t) u_\alpha ({\bf x}),
\pi ({\bf x}, t) = \sum_{\alpha}
\pi_\alpha (t) u_\alpha ({\bf x}).
\end{equation}
Then one obtains  the mode-decomposed
Hamiltonian
\begin{equation}
H(t) = \sum_{\alpha} H_\alpha (t) =
\sum_\alpha \left[
\frac{1}{2} \frac{1}{R^3 (t)}
\pi_\alpha^2 + \frac{1}{2} R^3 (t)
\left(m^2 + \xi {\bf R} +
\frac{{\bf k}^2}{R^2 (t)} \right) \phi_\alpha^2 \right].
\label{ham}
\end{equation}
The free massive scalar
field leads to the Hamiltonian which is a collection of
the time-dependent harmonic oscillators with
the time-dependent mass and frequency squared
\begin{equation}
M(t) = R^3 (t),
\omega_\alpha^2 (t) =
m^2 + \xi {\bf R} +
{\bf k}^2 /R^2 (t).
\end{equation}
We quantize the scalar field according to the
Schr\"{o}dinger-picture:
\begin{equation}
i \hbar \frac{\partial}{\partial t}
\Psi(\phi_\alpha,t)
= \hat{H} (t) \Psi (\phi_\alpha,t)
\label{Sch pic}.
\end{equation}
The Hamiltonian being mode-decomposed, the total wave function
for the scalar field is
\begin{equation}
\Psi (\phi_\alpha , t)
= \Pi_\alpha \Psi_{n_\alpha} (\phi_\alpha, t)
\end{equation}
a product of the wave function
$\Psi_{n_\alpha} (\phi_\alpha ,t) $ for each oscillator
$ H_\alpha (t) $.

\section{Fock Space out of Generalized Invariant}

In order to find the exact quantum states and
construct the Fock space for the Hamiltonian (\ref{ham}),
we shall use the generalized invariant method and
advance it to a field theoretical framework.
The most advantageous
point of the generalized invariant is the
substantiation and easiness in
constructing the Fock space of exact quantum states
and readiness in  applying to the density operator.
{}From the well-known result
of the generalized invariant called Lewis-Riesenfeld
invariant \cite{lewis},
the exact quantum state of the $\alpha$-th
oscillator's Schr\"{o}dinger
equation (\ref{Sch pic}) is given by
\begin{equation}
\Psi_{n_\alpha} (\phi_\alpha,t) = \exp \left(
- \int \left<\alpha, n_\alpha, t \right|
\frac{\partial}{\partial t} - \frac{1}{i \hbar}
\hat{H}_\alpha(t)
\left|\alpha, n_\alpha, t \right>dt \right)
\left|\alpha, n_\alpha, t \right>
\label{ex s}
\end{equation}
in terms of the eigenstates of the generalized invariant
\begin{equation}
\hat{I}_\alpha (t) \left|\alpha, n_\alpha, t \right>
=  \lambda_{\alpha n_\alpha} \left|\alpha, n_\alpha, t \right>,
\end{equation}
with a time-independent eigenvalue $\lambda_{\alpha n_\alpha}$,
which satisfies the invariant equation
\begin{equation}
\frac{\partial}{\partial t} \hat{I}_\alpha (t)
- \frac{i}{\hbar}
\left[\hat{I}_\alpha (t) , \hat{H}_\alpha (t) \right] = 0.
\label{inv eq}
\end{equation}

The most simplest way to construct the Fock
space is to find the
set of the first order fundamental invariants of the form
\begin{equation}
\hat{I}_{\alpha j} = \phi_{\alpha j} (t) \hat{\pi}_\alpha
- R^3 (t) \dot{\phi}_{\alpha j} (t) \hat{\phi}_\alpha
\label{cl sol}
\end{equation}
where $j$ runs for 1,2 and
\begin{equation}
\frac{\partial^2}{\partial t^2} \phi_{\alpha j}(t)
+ 3 \frac{\dot{R}(t)}{R(t)} \frac{\partial}{\partial t}
\phi_{\alpha j} (t) +
\left(m^2 + \xi {\bf R} +
\frac{{\bf k}^2}{R^2 (t)}\right)
\phi_{\alpha j} (t) = 0,
\end{equation}
are classical solutions of the $\alpha$-th oscillator.
We can make the first order invariants
satisfy the commutation relation
\begin{equation}
[\hat{I}_{\alpha 2} (t) , \hat{I}_{\beta 1} (t)] = \hbar
\delta_{\alpha \beta}
\label{com r}
\end{equation}
by imposing a condition on the integration constants
such that
\begin{equation}
R^3 (t) \left(\phi_{\alpha 1} (t) \dot{\phi}_{\alpha 2} (t)
- \phi_{\alpha 2} (t) \dot{\phi}_{\alpha 1} (t) \right) = i.
\label{int c}
\end{equation}
We see the group structure $SU^\infty(1,1)$
for the scalar field by forming
the second order independent invariants
\begin{equation}
\hat{I}^{(2)}_{\alpha jk} (t) = \frac{1}{2}
\left(\hat{I}_{\alpha j} (t) \hat{I}_{\alpha k} (t)
+ \hat{I}_{\alpha k} (t) \hat{I}_{\alpha j} (t) \right).
\end{equation}
and by redefining the basis
\begin{equation}
\hat{K}_{\alpha +} (t)
= \frac{1}{2\hbar} \hat{I}^{(2)}_{\alpha 11} (t),
\hat{K}_{\alpha -} (t)
= \frac{1}{2\hbar} \hat{I}^{(2)}_{\alpha 22} (t),
\hat{K}_{\alpha 0}  (t)
= \frac{1}{2\hbar} \hat{I}^{(2)}_{\alpha 12} (t).
\end{equation}
such that
\begin{equation}
\left[\hat{K}_{\alpha 0} (t) , \hat{K}_{\beta {\pm}} (t)
\right] = \pm \hat{K}_{\pm} (t) \delta_{\alpha \beta},
\left[\hat{K}_{\alpha +} (t) , \hat{K}_{\beta -} (t)
 \right] = -2 \hat{K}_{\alpha 0} (t) \delta_{\alpha \beta}.
\end{equation}
The strong point of the commutation relation (\ref{com r})
is the possibility to interpret the first
order invariants $\hat{I}_{\alpha 1} (t)$
and $\hat{I}_{\alpha 2}(t)$ as a kind of formal creation and
annihilation operators for a time-dependent oscillator
\begin{equation}
\hat{A}_{\alpha}^{\dagger}(t) = \hat{I}_{\alpha 1} (t),
\hat{A}_{\alpha} (t) = \hat{I}_{\alpha 2} (t),
\label{c-a op}
\end{equation}
by imposing further a condition on
complex classical solutions
\begin{equation}
{\rm Im} \left(
\frac{\dot{\phi}_{\alpha 1} (t)}{\phi_{\alpha 1} (t)}\right)
< 0,
\label{con}
\end{equation}
In an asymptotic region in which the mass and frequency
squared tend to constant values, the formal creation
and annihilation operators corresponds to the conventional
creation and annihilation operators up to some trivial
time-dependent phase factors.
{}From the group structure the operator $\hat{K}_{\alpha +}$
($\hat{K}_{\alpha -}$) raises (lowers)
the eigenstates of $\hat{K}_{\alpha 0}$ by two photons.

The ground state wave function
that should be annihilated
by $\hat{A}_\alpha (t)$ is given by
\begin{equation}
\Psi_{\alpha 0} (\phi_\alpha,t) = \left(\frac{1}{2 \pi \hbar
|\phi_{\alpha 1} (t)|^2}
\right)^{\frac{1}{4}}
\exp \left(
\frac{iR^3 (t) \dot{\phi}^*_{\alpha 1} (t)}
{2 \hbar \phi^*_{\alpha 1} (t)} \phi_\alpha^2 \right).
\end{equation}
The $n_\alpha$th wave function can
be obtained by applying the operator
$\hat{A}^{n_\alpha \dagger} (t)$
to the ground state wave function
\begin{equation}
\Psi_{n_\alpha} (\phi_\alpha,t) = \left(
\frac{1}{n_\alpha ! (\sqrt{\hbar})^{n_\alpha}}
\right)^{\frac{1}{2}}
\hat{A}_\alpha^{\dagger n_\alpha}(t)
\Psi_{\alpha 0} (\phi_\alpha,t).
\end{equation}

Now one finds relatively easily the dispersion
relations for each oscillator
\begin{eqnarray}
(\Delta \phi_\alpha )_{n_\alpha}^2 &=&
\phi_{\alpha 1} (t)
\phi^*_{\alpha 1} (t) \hbar (2n_\alpha + 1),
\\
(\Delta \pi_\alpha )_{n_\alpha}^2
&=& R^3 (t) \dot{\phi}_{\alpha 1} (t)
\dot{\phi}^*_{\alpha 1} (t) \hbar (2n_\alpha + 1),
\label{disperion}
\end{eqnarray}
and the expectation value of the Hamiltonian operator
\begin{equation}
\left< \hat{H}_\alpha (t) \right>_{n_\alpha}
= \frac{R^3 (t)}{2}\left(\dot{\phi}_{\alpha 1} (t)
\dot{\phi}^*_{\alpha 1} (t)
+ \left(m^2 + \xi {\bf R} +
\frac{{\bf k}^2}{R^2 (t)}\right)
\phi_{\alpha 1} (t) \phi^*_{\alpha 1} (t) \right)
\hbar (2n_\alpha + 1).
\end{equation}
One particular feature of the quantum states
found in this paper is that the expectation value
of the Hamiltonian, the time-time component of the
energy-momentum tensor, is proportional to
that of the classical one \cite{kim2}.

One also finds the Bogoliubov transformation
between the formal creation and annihilation
operators at two different times
\begin{eqnarray}
\hat{A}^{\dagger}_\alpha (t) &=& u_{\alpha 1}
\hat{A}^{\dagger}_\alpha (t_0)
+ u_{\alpha 2} \hat{A}_\alpha (t_0),
\nonumber\\
\hat{A}_\alpha (t) &=& u^*_{\alpha 2}
\hat{A}^{\dagger}_\alpha (t_0)
+ u^*_{\alpha 1} \hat{A}_\alpha (t_0),
\label{bo t}
\end{eqnarray}
where
\begin{eqnarray}
u_{\alpha 1} (t,t_0) &=& i R^3 (t)
\left(\dot{\phi}_{\alpha 1} (t)
\phi^*_{\alpha 1} (t_0)
- \phi_{\alpha 1} (t) \dot{\phi}^*_1 (t_0) \right),
\nonumber\\
u_{\alpha 2} (t,t_0) &=& i R^3 (t)
\left(\phi_{\alpha 1} (t) \dot{\phi}_{\alpha 1} (t_0)
- \dot{\phi}_{\alpha 1} (t) \phi_{\alpha 1} (t_0) \right).
\end{eqnarray}
By direct substitution one can show that
\begin{equation}
|u_{\alpha 1}(t,t_0)|^2 - |u_{\alpha 2}(t,t_0)|^2 = 1.
\end{equation}
Then there is a unitary transformation of the formal
creation and annihilation operators
at two different times
\begin{equation}
\hat{A}^{\dagger}_\alpha (t) = \hat{S}^{\dagger}_\alpha (t,t_0)
\hat{A}^{\dagger}_\alpha (t_0) \hat{S}_\alpha (t,t_0)
\end{equation}
where
\begin{equation}
\hat{S}_\alpha (t,t_0) =
\exp\left(i \theta_{\alpha 1} \hat{A}^{\dagger}_\alpha (t_0)
\hat{A}_\alpha (t_0) \right)
\exp \left(\frac{\nu_\alpha}{2}
e^{-i (\theta_{\alpha 1} - \theta_{\alpha 2})}
\hat{A}^{2\dagger}_\alpha (t_0) - {\rm h.c.} \right)
\end{equation}
is the squeeze operator, in which
\begin{equation}
u_{\alpha 1} (t,t_0) = \cosh \nu_\alpha e^{-i \theta_{\alpha 1}},
u_{\alpha 2} (t,t_0) = \sinh \nu_\alpha e^{-i \theta_{\alpha 2}}.
\label{sq op}
\end{equation}

We exploit the physical meaning of the
time-dependent vacuum state
$\left|0 , t \right>$
that is annihilated by all the formal
annihilation operators
\begin{equation}
\hat{A}_\alpha (t) \left| 0, t \right> = 0.
\end{equation}
Thus the vacuum state is the tensor product
of each ground state
\begin{equation}
\left|0 , t \right> = \Pi_\alpha
\otimes \left|\alpha, 0, t \right>.
\label{vac}
\end{equation}
{}From the Bogoliubov transformation (\ref{bo t})
it follows that the inner product of the
ground state of each mode at two different times
is given by
\begin{equation}
\left< \alpha, 0, t_0 | \alpha, 0, t \right>
= \left(\frac{1}{|u_\alpha (t, t_0)|} \right)^{\frac{1}{2}}
\end{equation}
and in the continuum limit
that of the field at two different times is given by
\begin{equation}
\left<0, t_0 | 0, t \right>
= \exp\left(- \frac{1}{2} \frac{l^3}{(2\pi)^3}
\int d {\bf k} \ln |u_{\alpha 1} (t, t_0 )|
\right).
\end{equation}
Noting that $|u_{\alpha 1} (t, t_0 )| \leq 1$,
in which the equality holds only when $t = t_0$,
this implies that in the infinite
volume limit the time-dependent vacuum state
is orthogonal each other at two different times
\begin{equation}
\left<0 , t_0 |0 , t \right> = 0.
\end{equation}
Thus we obtain explicitly the
infinitely many unitarily inequivalent
Fock representations\footnote{
It was pointed out by F. C. Khanna and
G. Vitiello that
the new vacuum (\ref{vac}) makes sense
in spite of the time-dependence
and may have a relation with the one
parameter-dependent vacuum in the
thermal field theory that
leads to the infinitely many unitarily
inequivalent Fock space. This point
suggests strongly that our functional
Schr\"{o}dinger-picture approach
may have a close connection with the thermal
field theory.}
\cite{vitiello}.
In this
sense our functional Schr\"{o}dinger-picture
approach to quantum fields in curved spacetimes
substantiates the Fock space used
in the thermal field theory.

The instability of the Fock space, that is
the unitary inequivalence of the
vacuum at any two different times originated
from
time-dependent metrics
leads to the particle creation
\begin{equation}
\left<0, t \right| \hat{N} (t) \left|
0, t \right>
= \sum_\alpha \left<\alpha, 0, t \right|
\hat{N}_\alpha (t) \left|\alpha,
0, t \right>
= \frac{l^3}{(2\pi)^3}
\int d{\bf k} |u_2 (k,t,t_0)|^2.
\end{equation}
Thus we recover in the functional Schr\"{o}dinger-picture
approach
the cosmological particle creation first
introduced by Parker \cite{parker}.

\section{Physical Application}

We apply the results in the previous sections to
a free massive scalar field in the deSitter
spacetime, a  spatially
flat Friedmann-Robertson-Walker universe
with the metric
\begin{equation}
ds^2 = -dt^2 + e^{2 H_0 t} d{\bf x}^2,
\end{equation}
where $H_0$ is the expansion rate driven by the
vacuum energy density.
Then one obtains the Hamiltonian
\begin{equation}
H (t) = \sum_{\alpha}
\left[\frac{1}{2} e^{-3H_0 t} \pi_\alpha^2
+ \frac{1}{2} e^{3H_0 t} \left(
m^2 + 12 \xi H_0^2 + {\bf k}^2 e^{-2H_0t}
 \right) \phi_\alpha^2\right].
\label{de ham}
\end{equation}
Each mode of the scalar field
has the mass and frequency squared
\begin{equation}
M(t) = e^{3H_0t},
\omega_\alpha^2 (t) =
m^2 + 12 \xi H_0^2 + {\bf k}^2 e^{-2H_0 t}.
\end{equation}
The classical equation of motion
\begin{equation}
\ddot{\phi}_\alpha (t) + 3H_0 \dot{\phi}_\alpha
+ \left(m^2 + 12 \xi H_0^2 + {\bf k}^2 e^{-2H_0 t}
\right)
\phi_\alpha (t) = 0,
\label{cl eq}
\end{equation}
has two independent complex solutions
\begin{equation}
\phi_{\alpha_1} (t) = \sqrt{\frac{\pi}{4H_0}}
e^{- \frac{3}{2} H_0 t}
H_\nu^{(1)} (z),
\phi_{\alpha_2} (t) = \sqrt{\frac{\pi}{4H_0}}
e^{- \frac{3}{2} H_0 t}
H_\nu^{(2)} (z),
\end{equation}
where $H_\nu^{(1)}$ and $H_\nu^{(2)}$ are the
Hankel functions of the first and second kind respectively,
and
\begin{equation}
\nu = \left(\frac{9}{4} - \frac{m^2 + 12\xi H_0^2}{H_0^2}
 \right)^{\frac{1}{2}},
z = \frac{k}{H_0}
e^{- H_0 t}.
\end{equation}

Following Sec. 3, we find explicitly
the particular invariants
in terms of classical solutions by
\begin{eqnarray}
\hat{I}_{\alpha 1} (t)  &=& \phi_{\alpha 1} (t)
\hat{\pi}_\alpha - e^{3H_0 t} \dot{\phi}_{\alpha 1} (t)
\hat{\phi}_\alpha,
\nonumber\\
\hat{I}_{\alpha 2} (t) &=& \phi_{\alpha 2} (t)
\hat{\pi}_\alpha - e^{3H_0} \dot{\phi}_{\alpha 2} (t)
\hat{\phi}_\alpha.
\label{inv}
\end{eqnarray}
The first order invariants act
as the time-dependent
creation and annihilation
operators
\begin{equation}
\hat{I}_{\alpha 1} (t) = \hat{A}_\alpha^\dagger (t),
\hat{I}_{\alpha 2} (t) = \hat{A}_\alpha (t),
\end{equation}
such that
\begin{equation}
\left[\hat{A}_\alpha (t), \hat{A}_\alpha^\dagger (t)
 \right]  = \hbar
\end{equation}
The ground state is annihilated by the annihilation operator
\begin{equation}
\hat{A}_\alpha (t) \left|\alpha, 0, t \right> = 0,
\label{gr st}
\end{equation}
and the $n_\alpha$th states are obtained by applying the creation
operators $n_\alpha$ times
\begin{equation}
\left|\alpha, n_\alpha, t \right> =
\frac{1}{\sqrt{n_\alpha!}}
\left(\hat{A}_\alpha^\dagger (t) \right)^{n_\alpha}
\left|\alpha, 0, t \right>.
\label{n st}
\end{equation}
It is worthy to note that the new vacuum state
(\ref{vac}) constructed here for the de Sitter
spacetime is nothing but the Bunch-Davies vacuum state\footnote{
This supports our earlier argument \cite{kim3}
that the ground state (\ref{vac})
of scalar field may play the role of the physical vacuum
extending the Minkowski, Bunch-Davies, and Hawking vacua.}
\cite{bunch}.

{}From the field operators at an equal time
\begin{equation}
\hat{\phi}({\bf x}, t) = \sum_{\alpha}
\hat{\phi}_\alpha (t) u_\alpha ({\bf x}),
\hat{\phi}({\bf y}, t) = \sum_{\beta}
\hat{\phi}_\beta (t) u_\beta ({\bf x}),
\end{equation}
we find the two-point function evaluated
with respect to the ground state
\begin{equation}
\left< \hat{\phi}({\bf x}, t)
\hat{\phi}({\bf y}, t) \right> =
\sum_{\alpha} u_\alpha ({\bf x})
u_\alpha ({\bf y}) \phi_{\alpha 2} (t)
\phi_{\alpha 2}^* (t).
\end{equation}
In the infinite volume limit, it takes the form
\begin{equation}
\left< \hat{\phi}({\bf x}, t)
\hat{\phi}({\bf y}, t) \right> =
\frac{1}{(2\pi)^3} \int d{\bf k}
e^{i {\bf k} \cdot ( {\bf x}  -
{\bf y})} |\phi( {\bf k}, t)|^2,
\end{equation}
where
\begin{equation}
\phi ({\bf k}, t) = \phi_{\alpha 2} (t) =
\sqrt{\frac{\pi}{4H_0}}
e^{- \frac{3}{2} H_0 t} H_\nu^{(1)} (z).
\end{equation}
One reads the correlation function
from the two-point function
\begin{eqnarray}
\left< \hat{\phi}^2({\bf x}, t) \right>
&=& \frac{1}{(2\pi)^3} \int d{\bf k}
|\phi( {\bf k}, t)|^2
\nonumber\\
&=&
\frac{H_0}{8\pi} e^{- H_0 t} \int_{0}^{\infty} dz
z \left(H_0^2 e^{2H_0 t} z^2 - \gamma^2
\right)^{\frac{1}{2}} |H_\nu^{(1)} (z)|^2,
\end{eqnarray}
which goes to infinity due to long
wavelength modes and can be regularized
using the covariant point-splitting method
to give proper
physical quantity \cite{bunch}.

One also finds explicitly the dispersion relations
between the $n_\alpha$th quantum state
\begin{eqnarray}
(\Delta \hat{\pi}_\alpha )_{n_\alpha}^2
&=& \frac{\pi H_0}{4}
e^{3 H_0 t} \left(\frac{3}{2} + z \right)^2
\left| z H_{\nu+ 1}^{(1)} (z) + \nu H_\nu^{(1)} (z) \right|^2
\hbar (2n_\alpha + 1),
\nonumber\\
(\Delta \hat{\phi}_\alpha )_n^2
&=& \frac{\pi}{4H_0}
e^{- 3 H_0 t} |H_\nu^{(1)} (z)|^2
\hbar (2n_\alpha + 1),
\end{eqnarray}
and the uncertainty relation
\begin{equation}
(\Delta \hat{\pi}_\alpha)_{n_\alpha}
(\Delta \hat{\phi}_\alpha)_{n_\alpha}
= \frac{\pi}{4} \left(\frac{3}{2} + z \right)
\left| H_\nu^{(1)} (z) \right|
\left|z H_{\nu +1}^{(1)} (z) + \nu H_\nu^{(1)} (z) \right|
\hbar (2n_\alpha+1).
\end{equation}
At an early stage of expansion, $ t \rightarrow
- \infty$ ($z \rightarrow \infty$),
the uncertainty relation takes the minimum value
$ (\Delta \hat{\pi}_\alpha)_{n_\alpha}
(\Delta \hat{\phi}_\alpha)_{n_\alpha}
\simeq \hbar (2n_\alpha + 1)/2$, which
means that the adiabatic theorem holds.
But at a later stage of expansion,
$ t \rightarrow \infty$ ($z \rightarrow 0$),
the uncertainty relation increases indefinitely
by an exponential factor $e^{2 \nu H_0 t}$.
The expectation value of the Hamiltonian operator
is given explicitly by
\begin{equation}
\left< \hat{H}_\alpha (t) \right>_{n_{\alpha}} =
\frac{\pi H_0}{8}
\Biggl[ \Bigl| \bigl(\frac{3}{2} + \nu \bigr)
H_\nu^{(1)} (z) + z H_{\nu + 1}^{(1)} (z) \Bigr|^2
+ \bigl( \frac{9}{4} - \nu^2 + z^2 \bigr)
|H_\nu^{(1)} (z) |^2 \Biggr] \hbar (2n_\alpha +1).
\end{equation}
$\left< \hat{H} (t) \right>$ diverges due to long wavelenth
modes and should be regularized to give the
correct back reaction $G_{0 0} = 8\pi \left<
\hat{H} (t) \right>$.
The number operator for each mode is
\begin{equation}
\hat{N}_\alpha (t) = \hat{A}_\alpha^\dagger (t)
\hat{A}_\alpha (t),
\end{equation}
and the number of particles created in each mode
during the evolution of field from an initial
time $t_0$ to a final time $t$ is
\begin{eqnarray}
\left<\alpha, 0, t \right| \hat{N}_\alpha (t)
\left|\alpha, 0, t_0 \right>
 &=&
\frac{\pi^2}{16} e^{3H_0 (t - t_0)}
\Biggl| \Bigl( \bigl( \frac{3}{2} + \nu \bigr)
H_\nu^{(1)} (z) + z H_{\nu + 1}^{(1)} (z)
\Bigr) H_\nu^{(1)} (z_0)
\nonumber\\
&&- \Bigl( \bigl( \frac{3}{2} + \nu \bigr)
H_\nu^{(1)} (z_0) + z_0 H_{\nu + 1}^{(1)} (z_0)
\Bigr) H_\nu^{(1)} (z)
\Biggr|^2.
\end{eqnarray}
One can show that the Bogoliubov transformation
and the number of particles created therefrom
is the same as those
obtained from the canonical approach based
on the classical solution space of the Klein-Gordon
equation by taking the correspondence
$ e^{ 3 H_0 t / 2}
\phi_{\alpha 1} (t) \leftrightarrow f (t)$
and interpreting as
$ e^{3 H_0 t / 2}
\phi_{\alpha 1} (t)  \leftrightarrow
f^{{\rm out}} (t),
e^{3 H_0 t / 2}
\phi_{\alpha 1} (t_0)  \leftrightarrow
f^{{\rm in}} (t_0)$.
A particularly useful method for the particle
creation and entropy production is
to introduce the squeeze operator (\ref{sq op}).
The amount of increase of the information theoretic
entropy is given explicitly by
\begin{equation}
\Delta S_{\rm ent} = \sum_\alpha 2 | \nu_\alpha |.
\end{equation}

In summary, we are able to construct
explicitly the Fock space of quantum fields
in curved spacetimes, whose vacuum state
provides the one parameter-dependent
unitarily inequivalent Fock representations
of the thermal field theory. Futhermore,
the two-point and correlation function,
the back reaction of the energy-momentum tensor,
and the particle creation number  are
explicitly calculable. The field theoretical method
develpoed in this talk can be applied equally to
the nonequilibrium
quantum field theory in curved spacetimes
derived from the Wheeler-DeWitt equation,
in which one is able to account self-consistently
both the quantum back reaction of matter fields
and the quantum gravitational corrections to
the field equation.

\vskip 5 mm
\centerline{\bf Acknowledgement}
This work was supported in part by
the Non-Directed Research Fund, Korea Research
Foundation, 1994, by the Korea Science and
Engineering Foundation under Grant
No 951-0207-056-2.

\end{document}